\documentclass[12pt]{article}
\usepackage{amsfonts}

\begin{document}

\centerline{\underline{\Large\textbf{Classical and Quantum Measurement Theory}}}\vspace{3ex}

\centerline{Peter Morgan\vspace{0.2ex}}
\centerline{peter.w.morgan@yale.edu\vspace{0.7ex}}
\centerline{Physics Department, Yale University,\vspace{-0.2ex}}
\centerline{ New Haven, CT 06520, USA.\vspace{1ex}}
\centerline{\today\vspace{4ex}}

\newcommand{\BLabstract}{1.9}
\newcommand{\BLmain}{1.9}
\newcommand{\BLsub}{1.65}
\newcommand{\BLbibliography}{1.35}
\newcommand{\XXnewpage}{\newpage}
\renewcommand{\BLabstract}{1.2}\renewcommand{\BLmain}{1.1}\renewcommand{\BLsub}{1}\renewcommand{\BLbibliography}{1}\renewcommand{\XXnewpage}{}

\renewcommand{\baselinestretch}{\BLabstract}\normalsize

{\large\noindent\textsf{{\bfseries Abstract:}
Classical and quantum measurement theories are usually held to be different because the algebra of classical measurements is commutative, however the Poisson bracket allows  noncommutativity to be added naturally. 
After we introduce noncommutativity into classical measurement theory, we can also add quantum noise, differentiated from thermal noise by Poincar\'e invariance. 
With these two changes, the extended classical and quantum measurement theories are equally capable, so we may speak of a single ``measurement theory''. 
The reconciliation of general relativity and quantum theory has been long delayed because classical and quantum systems have been thought to be very different, 
however this unification allows us to discuss a unified measurement theory for geometry in physics. 
\vspace{1ex}}}\newline
{\footnotesize\noindent\hspace*{0.5in}\sf Essay written for the Gravity Research Foundation 2022 Awards for Essays on Gravitation.}\vspace{3ex}

\XXnewpage
\renewcommand{\baselinestretch}{\BLmain}\normalsize

The classical measurement theory we are accustomed to is \emph{incomplete}: it cannot describe all the measurement results that can be described by quantum measurement theory.
Instead of asking how we can complete quantum mechanics so it can be more like classical mechanics, we should ask how we can complete classical measurement theory so it can be more like quantum measurement theory.

Some pairs of probability measures do not admit a joint probability measure, a circumstance that can be modeled by a noncommutative algebra of measurement operators but that \emph{cannot} be modeled by a commutative algebra of measurement operators, which is commonly known as \emph{measurement incompatibility}.
Classical measurement theory has long been straw-manned to allow only a commutative algebra of measurement operators, however we can also use the Poisson bracket to construct a noncommutative group algebra over a group of canonical and non-canonical transformations.
That larger algebra can be used as a classical measurement theory whenever measurement results have to be modeled by probability measures that do not admit joint probability measures.\vspace{-1ex}

{\renewcommand{\baselinestretch}{\BLsub}\sffamily
\setlength{\spaceskip}{0.35em plus 0.1em minus 0.21em}\setlength{\xspaceskip}{0.7em plus 0.1em minus 0.45em}
\begin{quote}
For a phase space $P$ and real-valued functions $u\,{:}\,P\,{\rightarrow}\,\mathbb{R}$, there is a commutative algebra of operators $\hat Y_u$ that act by multiplication, $\hat Y_u\hat Y_v\,{=}\,\hat Y_{uv}$, but the Poisson bracket allows us to construct a nontrivial Lie algebra of operators $\hat Y_u,\hat Z_u$ for which $[\hat Y_u,\hat Y_v]\,{=}\,0$, $[\hat Z_u,\hat Y_v]\,{=}\,\hat Y_{\{u,v\}}$ and $[\hat Z_u,\hat Z_v]\,{=}\,\hat Z_{\{u,v\}}$.
We can exponentiate the $\hat Z_u$ operators to give a group of canonical transformations, whereas exponentiating the $\hat Z_u$ \emph{and} the $\hat Y_u$ operators gives a larger group of canonical and non-canonical transformations.
We take the extended algebra of classical measurements to be the group algebra of that larger group.
We can also think of that extended algebra as an algebra of sums of compositions of $\hat Y_{u_1}\hat Y_{u_2}{\cdots}\hat Y_{u_m}\hat Z_{v_1}\hat Z_{v_2}{\cdots}\hat Z_{v_n}$ instead of the traditional choice, $\hat Y_{u_1}\hat Y_{u_2}{\cdots}\hat Y_{u_m}$.
\end{quote}}\vspace{-1ex}

We can quite straightforwardly construct concrete Hilbert space representations of these classical algebras by introducing a Gibbs thermal state, with the amplitude of the thermal noise determined by Boltzmann's constant and by the temperature.
To introduce Planck's constant, however, we have to introduce ideas from quantum field theory, where quantum fluctuations are Poincar\'e invariant with amplitude determined by Planck's constant, whereas thermal fluctuations are not invariant under boosts.
Such a difference can only be introduced in 1+1 or higher dimensions.
For a classical measurement theory, correspondingly, we can introduce both a Poincar\'e invariant noise, with amplitude determined by Planck's constant, and a thermal noise that is invariant under a subgroup of the Poincar\'e group that leaves invariant a single time-like 4-vector.

With these two classically natural extensions, which are both also well-motivated by empirical necessity, we can consider the classical and quantum measurement theories to be effectively a single measurement theory, however a conventional difference remains: the generator of time-like evolutions in quantum mechanics, the Hamiltonian operator, is required to be bounded below, whereas the generator of time-like evolutions in classical mechanics, the Liouville operator, is unbounded.
This difference makes quantum theory analytic in a very useful way that is not available for classical theory.
As a consequence, however, we have a somewhat unenviable choice:\vspace{-1.25ex}

{\renewcommand{\baselinestretch}{\BLsub}\sffamily
\begin{itemize}
\item Although joint probability measures associated with measurements at time-like separation can be modeled classically by operators that commute, the same joint probability measures are modeled quantum mechanically by operators that in general do not commute, which is made possible by using the L\"uders transformer (which is more familiarly known as collapse of the quantum state.)
\item Although incompatible probability measures can be modeled using functions of only quantum mechanical position and momentum operators, the same incompatible probability measures are modeled classically using functions of classical canonical transformations (which are made possible by the Poisson bracket), as well as of classical position and momentum operators.\vspace{-1ex}
\end{itemize}}
\noindent Either joint probability measures can be handled relatively easily or incompatible probability measures can be handled relatively easily.

The above is presented in a manner peculiar to the author in two recent articles\cite{AlgKoopman,CollapseProduct}, where references to a selection of other literature may also be found.
\vspace{4ex}

One approach to a unified measurement theory is to introduce \emph{an indexed set of measurement operators}, \mbox{$\{\hat M_i:i\in\mathcal{I}\}$,} with which we freely generate an associative $*$-algebra $\mathcal{A}$ over the complex field.
The index set $\mathcal{I}$ might be, for example, a finite set of integers, or, as for a Wightman field, a Schwartz space of complex-valued test functions defined on Minkowski space.
For the complex anti-linear adjoint of the $*$-algebra, $(\lambda\hat A\hat B)^\dagger=\lambda^*\hat B^\dagger\hat A^\dagger$ and $(\hat M_i)^\dagger=\hat M_{i^c}$, where the index set involution $i\mapsto i^c\in\mathcal{I}$ may be trivial or not.

{\setlength{\spaceskip}{0.5em plus 0.2em minus 0.3em}\setlength{\xspaceskip}{0.6em plus 0.2em minus 0.4em}
For such a $*$-algebra, a \emph{state} gives a complex-valued \emph{$N$-measurement expected value}\hspace{0.2em plus 0em minus 0.2em}%
\allowbreak\mbox{$\rho(\hat M_{i_1}\hat M_{i_2}{\cdots}\hat M_{i_N})$} for any ordered product $\hat M_{i_1}\hat M_{i_2}{\cdots}\hat M_{i_N}$, which extends to the whole of $\mathcal{A}$ by complex linearity,
\allowbreak\mbox{$\rho(\lambda\hat A+\mu\hat B)=\lambda\rho(\hat A)+\mu\rho(\hat B)$;}
is positive semi-definite, $\rho(\hat A^\dagger\hat A)\ge 0$;
is compatible with the adjoint, $\rho(\hat A^\dagger)=\rho(\hat A)^*$;
and is normalized for an identity operator $\hat 1$, $\rho(\hat 1)=1$.
The foundational example of a Gaussian state can be presented in generating function form as
\newcommand{\rme}{{\mathsf{e}}}%
\newcommand{\rmi}{{\mathsf{i}}}%
$$\displaystyle\rho(\rme^{\rmi\lambda_1\hat M_{i_1}}\rme^{\rmi\lambda_2\hat M_{i_2}}\cdots\rme^{\rmi\lambda_N\hat M_{i_N}})
=\exp\Bigl[-\sum_m\lambda_m^2(i_m^c,i_m)/2-\sum_{m<n}\lambda_m\lambda_n(i_m^c,i_n)\Bigr],$$
where $(i_m,i_n)$ is a positive semi-definite matrix for any finite number of indices $i_1, i_2, ..., i_N\,{\in}\,\mathcal{I}$.
$(i,j)$ and the involution $i^c$ determine the 2-measurement expected values $\rho(\hat M_i^\dagger\hat M_j)=(i,j)$ and $\rho(\hat M_i\hat M_j)=(i^c,j)$, which determine all $N$-measurement expected values for a Gaussian state.
For the commutator, we have $\rho(\hat A[\hat M_i,\hat M_j]\hat B)=[(i^c,j)-(j^c,i)]\rho(\hat A\hat B)$, so a Gaussian state gives us a representation of the Weyl-Heisenberg algebra, which will be trivial, however, if $(i^c,j)=(j^c,i)$, which we have been accustomed to calling ``classical''.}

A \emph{$G$-vacuum state}, for a symmetry group $G$ that acts on $\mathcal{I}$, is a state that is invariant under the action of any $g\in G$,
$\rho_G(\hat M_{g(i_1)}\hat M_{g(i_2)}{\cdots}\hat M_{g(i_n)})=\rho_G(\hat M_{i_1}\hat M_{i_2}{\cdots}\hat M_{i_n})$,
for which we also require compatibility with the involution, $g(i^c)=g(i)^c$.
For a Gaussian $G$-vacuum state, we require that $(g(i),g(j))=(i,j)$ for any $g\in G$.
For any $G$-vacuum state, we can construct a \emph{$G$-vacuum projection operator} $\hat V_G$, $G$-invariantly defined by $\rho_G(\hat A\hat V_G\hat B)=\rho_G(\hat A)\rho_G(\hat B)$.
With this, we can construct a $*$-algebra $\mathcal{A}_G$ that is freely generated by $\mathcal{A}$ and $\hat V_G$, which in general is noncommutative and for which we can prove that $\rho_G$ is also a state.
Starting from a $G$-vacuum state, we can construct many states that are not $G$-invariant, such as, for any operator $\hat X\in\mathcal{A}$, {\small$\displaystyle\rho_{GX}(\hat A)=\frac{\rho_G(\hat X^\dagger\hat A\hat X)}{\rho_G(\hat X^\dagger\hat X)}$}.
\vspace{4ex}

To focus on some of the commitments we might or might not  expect to make when constructing a measurement theory for geometry in physics, we consider the Wightman axioms, which construct a quantum field theory on Minkowski space, with the Poincar\'e group as the symmetry group $G$ of a $G$-vacuum state.
There are variations of the Wightman axioms in the literature, however there are three additional constraints that are central for the Bosonic case:\vspace{-1.25ex}

{\renewcommand{\baselinestretch}{\BLsub}\sffamily
\begin{enumerate}
\item any generator of a time-like translation has a spectrum that is bounded below;\label{BoundedBelow}\vspace{-1.5ex plus 0.5ex minus 0.5ex}
\item addition and scalar multiplication are defined for the index set of test functions and $\hat M_i$ is complex linear, \mbox{$\hat M_{\lambda i +\mu j}=\lambda\hat M_i +\mu\hat M_j$,} so that we can construct an operator-valued distribution $\hat M(x)$;\label{LinearityCondition}\vspace{-1.5ex plus 0.5ex minus 0.5ex}
\item if test functions $i$ and $j$ have space-like separated supports, \mbox{$\rho(\hat A\hat M_i\hat M_j\hat B)\,{=}\,\rho(\hat A\hat M_j\hat M_i\hat B)$}.\label{NoSignaling}
\end{enumerate}\vspace{-1ex}
}

\noindent We ought not necessarily to insist on {\sffamily(\ref{BoundedBelow})} for a measurement theory for geometry in physics, because it is not satisfied for a classical Liouvillian operator.
We also ought not necessarily to insist on the linearity {\sffamily(\ref{LinearityCondition})}, which is distinct from the linearity of the state, because \emph{this} linearity does not seem specially natural from a classical perspective, addition might not be available for index sets that are appropriate for a measurement theory for geometry in physics, and linearity may be part of why there are no interacting models of the Wightman axioms in 3+1-dimensions\cite{Fragment}.
The no-signaling condition {\sffamily(\ref{NoSignaling})} is satisfied for a Wightman field, however we can also insist on \emph{universal} no-signaling, so that $\mathcal{A}$ is commutative, in which case we would construct a noncommutative algebra $\mathcal{A}_G$ using the $G$-vacuum projection operator.

Methodologically, the above construction, intentionally following the mathematical structure of quantum field theory rather closely, is rather empiricist.
There is a focus on how we describe the measurements we perform, on the index set $\mathcal{I}$ and on the action of the invariance group $G$ on $\mathcal{I}$, instead of on what we measure.
There is, in the mathematical model suggested here, just one beable, the state, which models the recorded results of the measurements that we meticulously index.
What we might call ``what is'' is left monolithically undifferentiated.
This does not preclude a different mathematics in which the state is more differentiated ---as a field of what is instead of as a field of measurements--- however it is not easy to ensure mathematical control when a chaotic dynamics is applied to what will apparently have to be a non-differentiable mathematical structure.

For a measurement theory for geometry in physics we might use a group of causal structure preserving diffeomorphisms, $G_c$, say, however it is not clear what action of $G_c$ we should use on an index set $\mathcal{I}_c$.
If we first attempt to construct a Gaussian $G_c$-vacuum state, everything would also depend on what constructions we might use for $(i_m,i_n)$ as a $G$-invariant positive semi-definite matrix and for the involution $i^c$.
This final paragraph would be a very limited starting point if general relativity had been reconciled with quantum theory long ago, however the clarification of the relationship between classical and quantum measurement theories that is suggested here will hopefully be of some help.

\XXnewpage

{\renewcommand{\baselinestretch}{\BLbibliography}
\footnotesize
\sf\noindent Only references to the author's papers are included here. Please see the bibliographies in those papers, which refer to work on quantum measurement theory, quantum probability, algebraic quantum mechanics, local quantum physics, foundations of quantum mechanics, Koopman-von Neumann classical mechanics, signal analysis, philosophy of physics, and renormalization.\par}

\end{document}